# Entanglement in a 3-spin Heisenberg-XY chain with nearest-neighbor interactions, simulated in an NMR quantum simulator

K. Rama Koteswara Rao and Anil Kumar

*Centre for Quantum Information and Quantum Computation,*
*Department of Physics and NMR Research Centre,*
*Indian Institute of Science, Bangalore-560012.*

## Abstract

The evolution of entanglement in a 3-spin chain with nearest-neighbor Heisenberg-XY interactions for different initial states is investigated here. In an NMR experimental implementation, we generate multipartite entangled states starting from initial separable pseudo-pure states by simulating nearest-neighbor XY interactions in a 3-spin linear chain of nuclear spin qubits. For simulating XY interactions, we follow algebraic method of Zhang *et al.* [Phys. Rev. A 72, 012331 (2005)]. Bell state between end qubits has been generated by using only the unitary evolution of the XY Hamiltonian. For generating W-state and GHZ-state a single qubit rotation is applied on second and all the three qubits respectively after the unitary evolution of the XY Hamiltonian.

## I. Introduction

Entanglement is a potentially useful and key resource behind the power of quantum information science [1,2]. Quantum algorithms such as Deutsch-Jozsa, Grover's and Shor's algorithm out perform their classical counterparts [1]. In quantum communication protocols such as quantum cryptography and quantum teleportation, entangled states are distributed between distant locations to transfer quantum information [1,2]. The generation of entanglement between qubits has been realized by various experimental techniques [3-7]. It is well known, that any entangling two-qubit gate is universal for quantum computation, when assisted by single qubit operations [8]. In general, direct interactions between qubits are used to generate entanglement. However, in a large qubit system with short range interactions, direct interaction between qubits decays rapidly with distance. In such a scenario, it is important to consider generating entanglement between qubits which are not directly coupled.

The most direct method of generating entanglement between indirectly coupled qubits is based on SWAP operations. This method involves, generating entanglement between directly coupled qubits and then transfer this entanglement to the desired qubits by a series of SWAP operations between neighboring qubits. However, this method requires the ability to modulate the strength or nature of interactions between pairs of neighboring qubits in time. In other words, this method would require control fields on the system varying over the scale of the spacing between the qubits. In a large spin system, this method could be complicated as controlling individual qubits would be difficult in most of the current experimental techniques used in quantum information processing (QIP). Also, this method is more susceptible to errors because each external control can induce errors and the number of such controls increases with the number of qubits. One possible alternative to the above method is the use of flying or mobile qubits [9]. In this approach, entanglement is first generated between the flying qubit and one of the two qubits which are to be entangled. This flying qubit then traverses through a channel and reaches the desired location, where its state is mapped onto the second qubit [9]. This approach would also be complicated as it could involve interfacing between different physical systems such as stationary spins and photons or stationary and mobile spins. Under these circumstances it is very useful to have alternative to the above methods which uses minimal external control. Spin chains are one such possible alternative [9].

Recently spin chains are proposed as short distance communication channels for connecting registers of a quantum computer [10]. In that paper, quantum state transfer is achieved from one end of the spin chain to the other end purely through the natural dynamical evolution of Heisenberg interaction. Since then, many interesting protocols for quantum state transfer and entanglement transfer in spin chains have been proposed [9,11-14]. Some of these proposals have been verified experimentally by simulating the spin chains in a nuclear magnetic resonance quantum simulator [15-18]. The dynamics of entanglement in XY and Ising spin chains have also been studied [19-22]. Spin chains were also proposed for entanglement generation and distribution [9,23]. Quantum computation using permanently coupled spin chains has been suggested [9,24-26].

In this paper we focus on the dynamics of entanglement in a 3-spin linear chain with nearest-neighbor Heisenberg-XY interactions. We present an NMR experimental implementation of generation of multipartite entangled states (MES) starting from separable pseudo-pure states by simulating XY interactions in a linear chain of 3-nuclear spin qubits. Though, NMR experimental realizations of quantum state transfer (QST) have been reported [15-18], to the best of our knowledge this is the first experimental realization of generating MES in spin chains. In this work, Bell state between end qubits has been generated using only the unitary evolution of the XY Hamiltonian. For generating W-state and GHZ-state a single qubit rotation is applied on second and all the three qubits respectively after the unitary evolution of the XY Hamiltonian. For simulating XY interactions we followed the algebraic method of Zhang *et al.* [15]. Initial pseudo-pure states were also created using only nearest-neighbor interactions. Section II of this paper describes Heisenberg-XY interactions and the dynamics of entanglement in a 3-spin chain with nearest-neighbor XY interactions and section III describes NMR experimental implementation of preparation of pseudo-pure states and generation of MES using nearest-neighbor interactions.

## II. Heisenberg-XY interaction

**(i) System and Hamiltonian**

The Hamiltonian of a linear spin chain with nearest-neighbor (NN) Heisenberg-XY interactions is given by [11],

$$H_{NN}^{XY} = \sum_{i=1}^{N-1} \frac{J_i}{2} \left( \sigma_x^i \sigma_x^{i+1} + \sigma_y^i \sigma_y^{i+1} \right) = \sum_{i=1}^{N-1} J_i \left( \sigma_+^i \sigma_-^{i+1} + \sigma_-^i \sigma_+^{i+1} \right), \qquad (1)$$

Where $\sigma^i_{x/y}$ are the Pauli spin matrices, $\sigma^i_{+/-}$ are the raising and lowering operators, and $J_i$ is the coupling between spins $i$ and $i+1$. These Heisenberg-XY interactions can be used to transfer quantum states, distribute and generate entanglement over the spin chain with minimal external control of the qubits [9,11,12].

The Hamiltonian of a 3-spin linear chain with equal nearest-neighbor Heisenberg-XY interactions can be written as,

$$H^{XY}_{NN} = \frac{1}{2}J\left(\sigma^1_x\sigma^2_x + \sigma^1_y\sigma^2_y + \sigma^2_x\sigma^3_x + \sigma^2_y\sigma^3_y\right). \tag{2}$$

The unitary operator corresponding to the Hamiltonian given in equation (2) is given by,

$$U(t) = \exp(-iH^{XY}_{NN}t) = \exp\left(-i\frac{Jt}{2}\left(\sigma^1_x\sigma^2_x + \sigma^1_y\sigma^2_y + \sigma^2_x\sigma^3_x + \sigma^2_y\sigma^3_y\right)\right). \tag{3}$$

By using the commutation relation $\left[\left(\sigma^1_x\sigma^2_x + \sigma^2_y\sigma^3_y\right), \left(\sigma^1_y\sigma^2_y + \sigma^2_x\sigma^3_x\right)\right] = 0$, the above equation can be written as [15],

$$U(t) = \exp\left(-i\frac{Jt}{2}\left(\sigma^1_x\sigma^2_x + \sigma^2_y\sigma^3_y\right)\right)\exp\left(-i\frac{Jt}{2}\left(\sigma^1_y\sigma^2_y + \sigma^2_x\sigma^3_x\right)\right)$$
$$= \left[\cos(\varphi) I - \frac{i}{\sqrt{2}}\sin(\varphi)\left(\sigma^1_x\sigma^2_x + \sigma^2_y\sigma^3_y\right)\right]\left[\cos(\varphi) I - \frac{i}{\sqrt{2}}\sin(\varphi)\left(\sigma^1_y\sigma^2_y + \sigma^2_x\sigma^3_x\right)\right],$$

where $\varphi = Jt/\sqrt{2}$. The matrix form of the above unitary operator is given by [15],

$U(t) =$

$$\begin{bmatrix}
1 & 0 & 0 & 0 & 0 & 0 & 0 & 0 \\
0 & \cos^2\varphi & -\frac{i}{\sqrt{2}}\sin(2\varphi) & 0 & -\sin^2\varphi & 0 & 0 & 0 \\
0 & -\frac{i}{\sqrt{2}}\sin(2\varphi) & \cos(2\varphi) & 0 & -\frac{i}{\sqrt{2}}\sin(2\varphi) & 0 & 0 & 0 \\
0 & 0 & 0 & \cos^2\varphi & 0 & -\frac{i}{\sqrt{2}}\sin(2\varphi) & -\sin^2\varphi & 0 \\
0 & -\sin^2\varphi & -\frac{i}{\sqrt{2}}\sin(2\varphi) & 0 & \cos^2\varphi & 0 & 0 & 0 \\
0 & 0 & 0 & -\frac{i}{\sqrt{2}}\sin(2\varphi) & 0 & \cos(2\varphi) & -\frac{i}{\sqrt{2}}\sin(2\varphi) & 0 \\
0 & 0 & 0 & -\sin^2\varphi & 0 & -\frac{i}{\sqrt{2}}\sin(2\varphi) & \cos^2\varphi & 0 \\
0 & 0 & 0 & 0 & 0 & 0 & 0 & 1
\end{bmatrix},$$

(4)

where the states are ordered as $|000\rangle$, $|001\rangle$, $|010\rangle$, $|011\rangle$, $|100\rangle$, $|101\rangle$, $|110\rangle$, $|111\rangle$. By choosing appropriate $\varphi$, the operator $U(t)$ can perform various useful operations such as, state transfer and creation of entanglement. The latter is discussed below.

**(ii) Entanglement measures**

In a pure three qubit system, the following measures of entanglement are commonly used - the concurrence between two qubits, the concurrence between a qubit and rest of the system, and a measure of an intrinsic three partite entanglement [27].

The concurrence, which is a measure of the bipartite entanglement between two qubits, can be written as [28],

$$C_{ij} = \max\{\lambda_1 - \lambda_2 - \lambda_3 - \lambda_4, 0\}, \qquad (5)$$

where $\lambda_\alpha$ ($\alpha = 1, 2, 3, 4$) are the square roots of the eigen values of $\varrho_{ij}\widetilde{\varrho_{ij}}$ in decreasing order, $\varrho_{ij}$ is the reduced density matrix for the two qubits $i$ and $j$ and $\widetilde{\varrho_{ij}}$ is a spin flipped version of the density matrix $\varrho_{ij}$, i.e., $\widetilde{\varrho_{ij}} = \sigma_y \otimes \sigma_y \, \varrho_{ij}^* \, \sigma_y \otimes \sigma_y$. The concurrence is directly related to the entanglement of formation [28]. The minimal value of concurrence is zero, which means, the states of the two qubits are separable. For any value of concurrence greater than zero, the two qubits are entangled. The maximal value of concurrence is one, which means the states of the two qubits are maximally entangled and are locally unitary equivalent to Bell states [28].

In a tripartite system, the concurrence between a qubit *i* and rest of the system (*jk*) can be written as [29],

$$C_{i(jk)} = [2(1 - Tr(\rho_i^2))]^{1/2}, \qquad (6)$$

where $\rho_i$ is the reduced density operator for the qubit *i*.

Finally, the intrinsic three qubit entanglement, which is defined for pure states can be written as [27],

$$C_{ijk} = C_{i(jk)}^2 - C_{ij}^2 - C_{ik}^2. \qquad (7)$$

For tripartite entangled GHZ and W-states, the intrinsic three qubit entanglement $C_{ijk}$ is one and zero respectively. On the other hand, the concurrence (bipartite entanglement) between pairs of qubits is zero for GHZ state and 2/3 for W-state.

**(iii) Entanglement dynamics**

We first study the evolution of entanglement in a three spin chain with nearest-neighbor XY interactions, starting from the initial separable pure states. For the ferromagnetic pure states $|000\rangle$ and $|111\rangle$, no entanglement is generated during the unitary evolution, as these are the eigen states of the XY-Hamiltonian. For the states $|010\rangle$ and $|101\rangle$, where the state of the middle qubit is reversed from the state of the end qubits, the dynamics of entanglement during the unitary evolution is shown in Fig. 1(a). The concurrence between qubits *1* and *3*, $C_{13}$, reaches its maximum value of 1 at points $\varphi = (2n + 1)\pi/4$, where $n = 0, 1, 2 \ldots$.

$$U\left(\frac{\pi}{2\sqrt{2}J}\right)|010\rangle = -\frac{i}{\sqrt{2}}(|001\rangle + |100\rangle), \qquad (8)$$

$$U\left(\frac{\pi}{2\sqrt{2}J}\right)|101\rangle = -\frac{i}{\sqrt{2}}(|011\rangle + |110\rangle). \tag{9}$$

In the above equations the 1<sup>st</sup> and 3<sup>rd</sup> qubits are in the Bell state $\left[\frac{1}{\sqrt{2}}(|01\rangle + |10\rangle)\right]$ and the second qubit is in the state $|0\rangle$ or $|1\rangle$ depending upon the initial state. Thus, a Bell state between end qubits can be generated starting from the initial separable states using only the natural dynamical evolution of the nearest-neighbor XY Hamiltonian. The concurrence between qubits *1* and *2*, $C_{12}$, reaches a maximum value of $1/\sqrt{2}$ at points $\varphi = (2n+1)\pi/8$, where $n = 0, 1, 2 \ldots$ . No intrinsic tripartite entanglement $C_{123}$ is generated during the unitary evolution for the initial states $|010\rangle$ and $|101\rangle$. When the initial state $|101\rangle$ is evolved under $U(t)$, for a time $t = (\tan^{-1}\sqrt{2})/\sqrt{2}J$ ($\varphi = \pi/6.577$), we obtain,

$$U\left(\frac{\tan^{-1}\sqrt{2}}{\sqrt{2}J}\right)|101\rangle = \frac{1}{\sqrt{3}}(|101\rangle - i|011\rangle - i|110\rangle). \tag{10}$$

After applying a $\pi/2$ phase gate on the second qubit, the above state becomes $\left[\frac{1}{\sqrt{3}}(|101\rangle + |011\rangle + |110\rangle)\right]$, which is a W-state. The unitary evolution of the initial equal superposition state $\left[\frac{1}{\sqrt{2}}(|0\rangle + |1\rangle) \otimes \frac{1}{\sqrt{2}}(|0\rangle + |1\rangle) \otimes \frac{1}{\sqrt{2}}(|0\rangle + |1\rangle)\right]$ is shown in Fig. 1(b). The concurrence between qubits *1* and *3*, $C_{13}$, is zero during the unitary evolution for this initial state. The intrinsic three partite entanglement $C_{123}$ reaches its maximum value of 1 at $\varphi = \pi/2$. This corresponds to the state $\left[\frac{1}{\sqrt{8}}(|000\rangle - |001\rangle - |010\rangle - |011\rangle - |100\rangle - |101\rangle - |110\rangle + |111\rangle)\right]$. A $\pi/2$ X-rotation of all the three qubits converts this state into a state $\left[\frac{1}{\sqrt{2}}(|000\rangle + |111\rangle)\right]$, which is a GHZ state. Thus, tripartite entangled states, W and GHZ states can be generated from the initial separable states by applying a single qubit rotation of the middle qubit and all the qubits respectively, after the unitary evolution of the nearest-neighbor XY Hamiltonian. The evolution of entanglement for the initial states $|001\rangle$, $|110\rangle$ and $|011\rangle$, $|100\rangle$ is also shown in the Figs. 1(c) and 1(d). The evolution of $C_{1(23)}$ reaches its maximum value of 1 at different values of $\varphi$ for different initial states as shown in Fig. 1.

The following section contains experimental NMR implementation of the generation of the entangled states discussed above.

## III. Experimental Implementation

The spin system chosen for the experimental implementation is $^{13}$C labeled dibromo-fluoro-methane ($^{13}$CHFBr$_2$), where $^1$H, $^{13}$C and $^{19}$F act as three qubits. The chemical structure of the molecule, the J-couplings and the equilibrium spectra are shown in Fig. 2. The molecule was synthesized in our laboratory following reported synthesis in literature [30,31]. This unique molecule has very large one-bond J-couplings between nearest neighbors, resulting very small gate times (~ 3 to 5 msec), much less than the coherence times of each qubit. The experiments have been carried out at a temperature of 300K in 11.7 Tesla magnetic fields on an AV 500 spectrometer using a triple resonance QXI-probe. The $^1$H, $^{13}$C and $^{19}$F resonance frequencies at this field are 500 MHz, 125 MHz and 470 MHz respectively. The scalar couplings between spins are $J_{HC} = 224.5\ Hz$, $J_{CF} = -310.9\ Hz$ and $J_{HF} = 49.7\ Hz$. The measured relaxation times are T$_1$(H)=6.7 sec, T$_1$(C)=1.9 sec, T$_1$(F)=4.0 sec, T$_2$(H)=1400 msec, T$_2$(C)=710 msec and T$_2$(F)=700 msec.

In the three qubit spin chain we label, qubit 1=$^1$H, qubit 2=$^{13}$C and qubit 3=$^{19}$F. In the molecule $^{13}$CHFBr$_2$, all the three couplings ($J_{12}$, $J_{23}$, $J_{13}$) are present. However, the coupling '$J_{13}$' has been refocused during the total time of all the experiments, thus retaining only the nearest-neighbor couplings.

In the triply rotating frame-resonance (with ON resonance observation of all the three spins), the NMR Hamiltonian of the three qubit system, where the interaction between qubits is restricted to the nearest neighbors only, is given by,

$$H_{NMR} = \frac{\pi}{2} J_{12} \sigma_z^1 \sigma_z^2 + \frac{\pi}{2} J_{23} \sigma_z^2 \sigma_z^3. \tag{11}$$

### (i) Preparation of Pseudo-Pure states (PPS)

In liquid state room temperature NMR, it is well known, that since preparation of a pure state requires extreme conditions, a pseudo pure state is prepared that mimics a pure state [4,32]. The equilibrium deviation density matrix under high temperature and high field approximation, is in a highly mixed state, which can be represented by [33],

$$\Delta \rho_{eq} \propto \gamma_1 \sigma_z^1 + \gamma_2 \sigma_z^2 + \gamma_3 \sigma_z^3 = \gamma_1(\sigma_z^1 + 0.25\sigma_z^2 + 0.94\sigma_z^3), \qquad (12)$$

where $\gamma_1$, $\gamma_2$ and $\gamma_3$ are the gyromagnetic ratios of $^1$H, $^{13}$C and $^{19}$F respectively. The initial pseudo-pure state $|000\rangle$ was prepared from the equilibrium density matrix $\rho_{eq}$ by the spatial averaging method [34]. The $|000\rangle$ PPS is obtained in terms of product of operators as [31],

$$\rho_{|000\rangle}^{PPS} = \frac{\gamma_1}{4}(\sigma_z^1 + \sigma_z^2 + \sigma_z^3 + 2\sigma_z^1\sigma_z^2 + 2\sigma_z^2\sigma_z^3 + 2\sigma_z^1\sigma_z^3 + 4\sigma_z^1\sigma_z^2\sigma_z^3). \qquad (13)$$

The radio frequency and magnetic field gradient pulses, shown in Fig. 3(a) transform the spin system from the equilibrium state to $|000\rangle$ pseudo-pure state, using only the nearest-neighbor couplings. It may be noted that earlier we have created PPS in this spin system using all the three couplings [31]. However, in the present case we have used only the nearest-neighbor couplings necessitating an additional evolution period. The other PPSs can easily be created from $|000\rangle$ PPS by spin selective $(\pi)_x$ pulses, for example $|000\rangle \xrightarrow{\pi_x^1} |100\rangle \xrightarrow{\pi_x^2} |110\rangle \xrightarrow{\pi_x^3} |111\rangle$. The real part of the experimentally determined density matrix for the $|010\rangle$ PPS is shown in Fig. 4. The fidelity of the PPSs was calculated by the formula [35],

$$F(\rho_{th}, \rho_{ex}) = \frac{Tr(\rho_{th}\rho_{ex})}{\sqrt{Tr(\rho_{th}^2) \, Tr(\rho_{ex}^2)}}, \qquad (14)$$

where, $\rho_{th}$ is theoretically expected density matrix and $\rho_{ex}$ is experimentally obtained density matrix. All the PPSs created by above method have a fidelity of $\approx 0.99$.

**(ii) Generation of entangled states**

The unitary operator corresponding to nearest-neighbor Heisenberg-XY interaction in a 3-spin chain (equation (3)) can be written as,

$$U(\sqrt{2}\varphi/J) = \exp\left(-\frac{i\varphi}{\sqrt{2}}(\sigma_x^1\sigma_x^2 + \sigma_y^1\sigma_y^2 + \sigma_x^2\sigma_x^3 + \sigma_y^2\sigma_y^3)\right). \tag{15}$$

In the weakly coupled spin system of NMR, the Heisenberg-XY interactions are not present directly. The interactions present are the ZZ type interactions. Following the method of Zhang *et al.*, the unitary operator in the equation (15) has been simulated using ZZ-interactions and radio-frequency pulses [15],

$$\begin{aligned}
U(\sqrt{2}\varphi/J) &= e^{-i(\pi/4)\sigma_y^1}\, e^{i(\pi/4)\sigma_x^3}\left(e^{-i(\pi/8)\sigma_z^1\sigma_z^2\sigma_z^3}\right) e^{-i(\pi/4)\sigma_y^2}\left(e^{-i(\varphi)\sigma_z^1\sigma_z^2}\right) e^{i(\pi/4)\sigma_y^2} \\
&\quad \times \left(e^{i(\pi/8)\sigma_z^1\sigma_z^2\sigma_z^3}\right) e^{i(\pi/4)\sigma_y^1}\, e^{-i(\pi/4)\sigma_x^3}\, e^{i(\pi/4)\sigma_x^1}\, e^{-i(\pi/4)\sigma_y^3}\left(e^{-i(\pi/8)\sigma_z^1\sigma_z^2\sigma_z^3}\right) \\
&\quad \times e^{-i(\pi/4)\sigma_y^2}\left(e^{-i(\varphi)\sigma_z^2\sigma_z^3}\right) e^{i(\pi/4)\sigma_y^2}\left(e^{i(\pi/8)\sigma_z^1\sigma_z^2\sigma_z^3}\right) e^{-i(\pi/4)\sigma_x^1}\, e^{i(\pi/4)\sigma_y^3}.
\end{aligned} \tag{16}$$

The 3-spin interaction terms in equation (16) can be generated from 2-spin interaction terms and radio frequency pulses, following the method of Tseng *et al.* [36] is obtained as,

$$\begin{aligned}
e^{-i(\pi/8)\sigma_z^1\sigma_z^2\sigma_z^3} &= e^{-i(\pi/4)\sigma_x^2}\, e^{-i(\pi/4)\sigma_z^1\sigma_z^2}\, e^{-i(\pi/4)\sigma_y^2}\, e^{-i(\pi/8)\sigma_z^2\sigma_z^3} \\
&\quad \times e^{-i(\pi/4)\sigma_y^2}\, e^{-i(\pi/4)\sigma_z^1\sigma_z^2}\, e^{i(\pi/2)\sigma_y^2}\, e^{i(\pi/4)\sigma_x^2}.
\end{aligned} \tag{17}$$

Single spin rotations, for example $e^{-i(\pi/4)\sigma_x^2}$ can be implemented by a $\pi/2$ pulse on spin *2* with phase *x*. The 2-spin interaction terms, for instance $U^{12} = e^{-i(\phi)\sigma_z^1\sigma_z^2}$ can be realized by the pulse sequence $\tau/2 - [\pi]_3 - \tau/2 - [\pi]_3$, where $\tau = 2\phi/(\pi J_{12})$. Substituting equation (17) into equation (16) and after some simplifications, the NMR pulse sequence, corresponding to the unitary operator of equation (16) for the generation of entangled states, is given in Fig. 3(b).

To quantitatively evaluate the experimental results, we calculate the attenuated correlation [37], given by,

$$c(\rho_{ex}) = \frac{Tr(\rho_{th}\rho_{ex})}{Tr(\rho_{th}\rho_{th})}, \qquad (18)$$

where, $\rho_{th}$ is the theoretical density matrix obtained after applying ideal transformations to the experimental initial density matrix, and $\rho_{ex}$ is the experimentally determined final density matrix.

*(a) Bell state on end qubits*

The NMR pulse sequence, shown in Fig. 3(b), with $d_1 = 1/2J_{23}$ and $d_2 = 1/2J_{12}$, which corresponds to $\varphi = \pi/4$, was used to implement the unitary operator of equations (8) and (9). This sequence was applied on initial PPSs $|010\rangle$ and $|101\rangle$ to yield Bell state on end qubits with middle qubit being in state $|0\rangle$ and $|1\rangle$ respectively. Complete quantum state tomography [38] of the final density matrix was performed and the results are shown in Fig. 5(a) and 5(b). These results confirm the creation of Bell state on end qubits $\left[-\frac{i}{\sqrt{2}}(|001\rangle + |100\rangle)\right]$ and $\left[-\frac{i}{\sqrt{2}}(|011\rangle + |110\rangle)\right]$ with a correlation (c) of $\approx 0.86$ and $\approx 0.88$ respectively.

*(b) W-state*

The application of the NMR pulse sequence, shown in Fig. 3(b), with $d_1 = 0.3041/J_{23}$ and $d_2 = 0.3041/J_{12}$, which corresponds to $\varphi = \pi/6.577$, on the initial PPS $|101\rangle$, produces the state $\left[\frac{1}{\sqrt{3}}(|101\rangle - i|011\rangle - i|110\rangle)\right]$. The application of a $\pi/2$ phase gate on the second qubit yields the desired W-state $\left[\frac{1}{\sqrt{3}}(|101\rangle + |011\rangle + |110\rangle)\right]$. The phase gate was implemented by a $[\pi/2]$ Z-rotation of the second qubit which in turn was implemented by the composite rotation $[\pi/2]_Y - [\pi/2]_X - [\pi/2]_{-Y}$ [33]. Quantum state tomography of the final density matrix was performed to confirm the creation of the W-state and is shown in Fig. 5(c). The experimental

Pauli set $\vec{P}$ [39], whose 64 elements are the expectation values of the three-qubit Pauli operators, $\langle LMR \rangle$, where $L, M, R \in \{I, \sigma_x, \sigma_y, \sigma_z\}$, for the W-state is also shown in Fig. 6. The correlation for the W-state was found to be $c(\rho_{ex}^W) \approx 0.89$.

*(c) GHZ-state*

The equal superposition state $\left[\frac{1}{\sqrt{2}}(|0\rangle + |1\rangle) \otimes \frac{1}{\sqrt{2}}(|0\rangle + |1\rangle) \otimes \frac{1}{\sqrt{2}}(|0\rangle + |1\rangle)\right]$ was created from the $|000\rangle$ pseudo-pure state by applying a Hadamard gate on all the three qubits. The Hadamard gate was implemented by a $\pi/2$ Y-rotation. The application of the NMR pulse sequence, shown in Fig. 2(b) with $d_1 = 1/J_{23}$ and $d_2 = 1/J_{12}$, which corresponds to $\varphi = \pi/2$, on the equal superposition state, produces the state $\left[\frac{1}{\sqrt{8}}(|000\rangle - |001\rangle - |010\rangle - |011\rangle - |100\rangle - |101\rangle - |110\rangle + |111\rangle)\right]$. This state was then converted into the desired GHZ-state $\left[\frac{1}{\sqrt{2}}(|000\rangle + |111\rangle)\right]$ by a $\pi/2$ X-rotation of all the three qubits. To confirm the creation of GHZ-state, quantum state tomography of the final density matrix was performed and the results are shown in the Fig. 5(d). The correlation for the GHZ state was found to be $c(\rho_{ex}^{GHZ}) \approx 0.81$.

We also measured experimentally the evolution of the two-body correlation $\langle \sigma_x^1 \sigma_x^3 \rangle$ of the end qubits under XY Hamiltonian at different times, from $\varphi = 0$ to $\pi$, for the initial pseudo-pure state $|010\rangle$. The result of this measurement is shown in the Fig. 7. As expected, $\langle \sigma_x^1 \sigma_x^3 \rangle$ reaches maximum at $\varphi = \pi/4$ and $3\pi/4$. Comparing Fig. 7 with Fig. 1(a), the two-body correlation $\langle \sigma_x^1 \sigma_x^3 \rangle$ has a similar behavior to the concurrence ($C_{13}$) between end qubits.

The loss of correlation between experimental and theoretical density matrices is largely due to decoherence and pulse imperfections arising from rf field inhomogeneity. To quantitatively evaluate the error due to decoherence, we simulated the experiments, shown in Fig. 5 with the measured T1 and T2 using the decoherence model of Vandersypen *et al.* [40]. The error due to decoherence was estimated to be $\approx 6\%$, that is, $c_{dec} \approx 0.94$.

## IV. Conclusion

This paper demonstrates experimentally, the generation of multipartite entangled states in a linear chain of 3-nuclear spin qubits, by simulating the nearest-neighbor Heisenberg-XY interactions in an NMR set up. While extension to larger spin chains would be interesting, simulating XY interactions using ZZ interactions and single qubit rotations might need very long pulse sequences. The dipolar coupled spin systems, where the XY interactions are naturally present may well suit the above purpose. The simulation can be extended to spin chains with other type of interactions, for example, Ising or Heisenberg XYZ.

## Acknowledgement

The use of AV500 FTNMR spectrometer of NMR Research Centre funded by Department of Science and Technology (DST), New Delhi, is gratefully acknowledged. We also thank DST for funding the Centre for Quantum Information and Quantum Computation at Indian Institute of Science, Bangalore. Discussions with Prof. Apoorva Patel and V. S. Manu are gratefully acknowledged.

## References

[1] M. A. Nielson, and I. L. Chuang, Quantum Computation and Quantum Information, Cambride University Press, (2002).

[2] R. Horodecki, P. Horodecki, M. Horodecki, and K. Horodecki, Rev. Mod. Phys. 81, 865 (2009).

[3] E. Hagley, X. Maitre, G. Nogues, C. Wunderlich, M. Brune, J. M. Raimond, and S.Haroche, Phys. Rev. Lett. 79, 1 (1997).

[4] N. A. Gershenfeld, and I. L. Chuang, Science 275, 350 (1997).

[5] Q. A. Turchette, C. S. Wood, B. E. King, C. J. Myatt, D. Leibfried, W. M. Itano, C. Monroe, and D. J. Wineland, Phys. Rev. Lett. 81, 3631 (1998).

[6] M. S. Kim, J. Lee, D. Ahn, and P. L. Knight, Phys. Rev. A 65, 040101 (2002).

[7] L. M. Duan, M. D. Lukin, J. I. Cirac, and P. Zollar, Nature 414, 413 (2001).

[8] M. J. Bremner, C. M. Dawson, J. L. Dodd, A. Gilchrist, A. W. Harrow, D. Mortimer, M. A.

Nielson, and T. J. Osborne, Phys. Rev. Lett. 89, 247902 (2002).

[9] S. Bose, Contemp. Phys. 48, 13 (2007).

[10] S. Bose, Phys. Rev. Lett. 91, 207901 (2003).

[11] M. Christandl, N. Datta, A. Ekert, and A. J. Landahl, Phys. Rev. Lett. 92, 187902 (2004).

[12] M. Christandl, N. Datta, T. C. Dorlas, A. Ekert, A. Kay, and A. J. Landahl, Phys. Rev. A 71, 032312 (2005).

[13] D. Burgarth, V. Giovannetti, and S. Bose, Phys. Rev. A 75, 062327 (2007).

[14] C. Di Franco, M. Paternostro, and M. S. Kim, Phys. Rev. Lett. 101, 230502 (2008).

[15] J. Zhang, G. L. Long, W. Zhang, Z. Deng, W. Liu, and Z. Lu, Phys. Rev. A 72, 012331 (2005).

[16] J. Zhang, N. Rajendran, X. Peng, and D. Suter, Phys. Rev. A 76, 012317 (2007).

[17] J. Zhang, M. Ditty, D. Burgarth, C. A. Ryan, C. M. Chandrashekar, M. Laforest, O. Moussa, J. Baugh, and R. Laflamme, Phys. Rev. A 80, 012316 (2009).

[18] G. A. Alvarez, M. Mishkovsky, E. P. Danieli, P. R. Levstein, H. M. Pastawski, and L. Frydman, Phys. Rev. A 81, 060302(R) (2010).

[19] V. Subrahmanyam, Phys. Rev. A 69, 034304 (2004).

[20] L. Amico, A. Osterloh, F. Plastina, R. Fazio, and G. M. Palma, Phys. Rev. A 69, 022304 (2004).

[21] F. Plastina, and T. J. G. Apollaro, Phys. Rev. Lett. 99, 177210 (2007).

[22] G. B. Furman, V. M. Meerovich, and V. L. Sokolovsky, Phys. Rev. A 77, 062330 (2008).

[23] M.-H. Yung, and S. Bose, Phys. Rev. A 71, 032310 (2005).

[24] S. C. Benjamin, and S. Bose, Phys. Rev. Lett. 90, 247901 (2003).

[25] M.-H. Yung, S. C. Benjamin, and S. Bose, Phys. Rev. Lett. 96, 220501 (2006).

[26] A. Kay, and M. Ericsson, New J. Phys. 7, 143 (2005).

[27] V. Coffman, J. Kundu, and W. K. Wootters, Phys. Rev. A 61, 052306 (2000).

[28] W. K. Wootters, Phys. Rev. Lett. 80, 2245 (1998).

[29] P. Rungta and C. M. Caves, Phys. Rev. A 67, 012307 (2003).

[30] L. M. K. Vandersypen, 'Experimental Quantum Computation with Nuclear Spins in Liquid Solution', Doctoral Thesis (2001) pp 167.

[31] Avik Mitra, K. Sivapriya, and Anil Kumar, J. Magn. Res. 187, 306 (2007).

[32] D. G. Cory, A. F. Fahmy, and T. F. Havel, Proc. Natl. Acad. Sci. USA 94, 1634 (1997) .

[33] R. R. Ernst, G. Bodenhausen, and A. Wokaun, Principles of Nuclear Magnetic Resonance in one and two dimensions, Oxford University Press (1990).

[34] D. G. Cory, M. D. Price, and T. F. Havel, Physica D 120, 82 (1998).

[35] X. Peng, J. Zhang, J. Du, and D. Suter, Phys. Rev. A 81, 042327 (2010).

[36] C. H. Tseng, S. Somaroo, Y. Sharf, E. Knill, R. Laflamme, T. F. Havel, and D. G. Cory, Phys. Rev. A 61, 012302 (1999).

[37] G. Teklemariam, E. M. Fortunato, M. A. Pravia, T. F. Havel, and D. G. Cory, Phys. Rev. Lett. 86, 5845 (2001).

[38] Avik Mitra, Arindam Ghosh, Ranabir Das, Apoorva Patel, and Anil Kumar, J. Magn. Res. 177, 285 (2005).

[39] J. M. Chow, L. DiCarlo, J. M. Gambetta, A. Nunnenkamp, L. S. Bishop, L.Frunzio, M. H. Devoret, S. M. Girvin, and R. J. Schoelkopf, Phys. Rev. A 81, 062325 (2010).

[40] L. M. K. Vandersypen, M. Steffen, G. Breyta, C. S. Yannoni, M. H. Sherwood and I. L. Chuang, Nature (London) 414, 883 (2001).

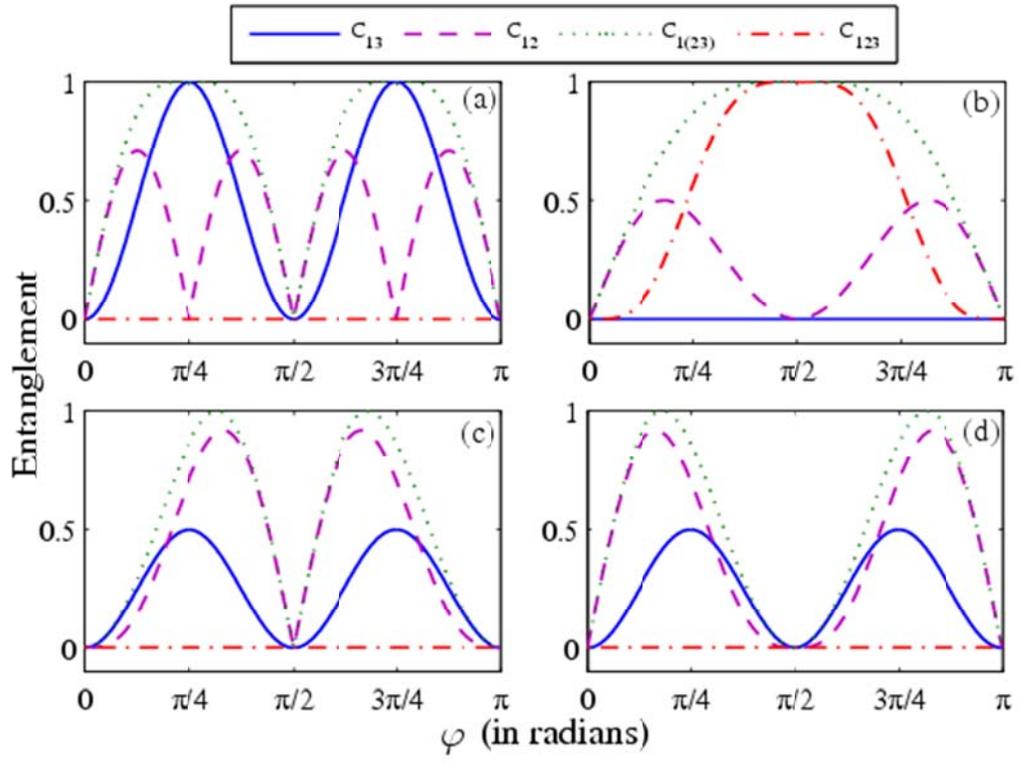

**FIG. 1.**

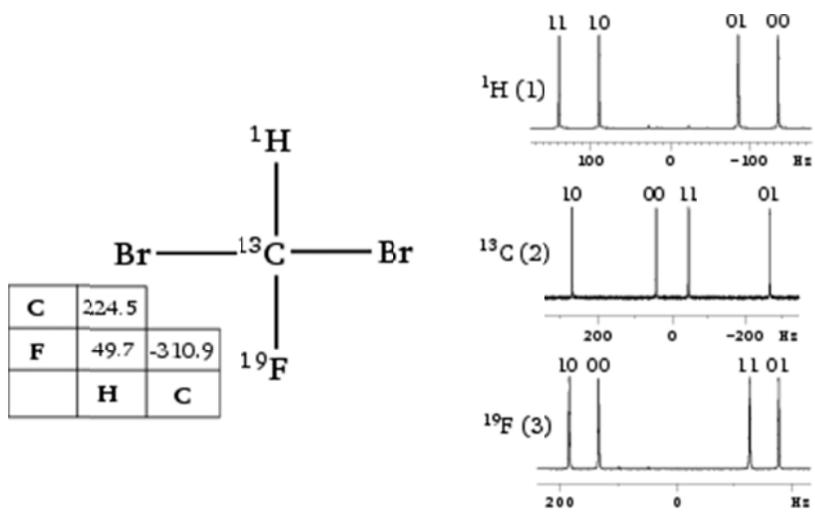

**FIG. 2.**

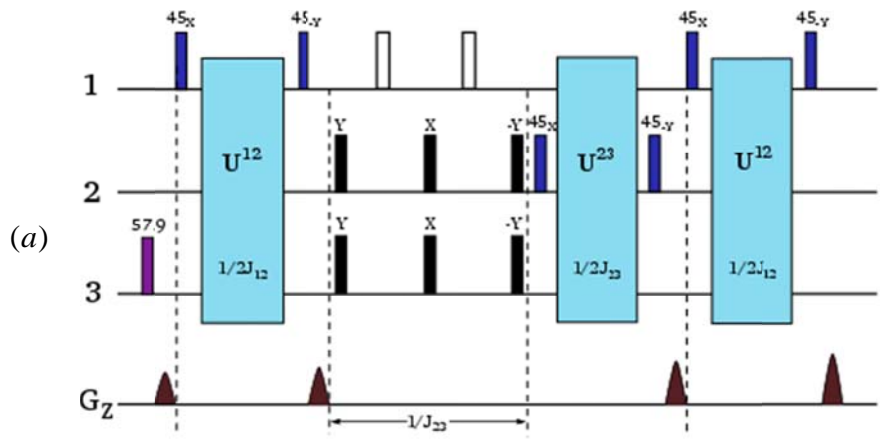

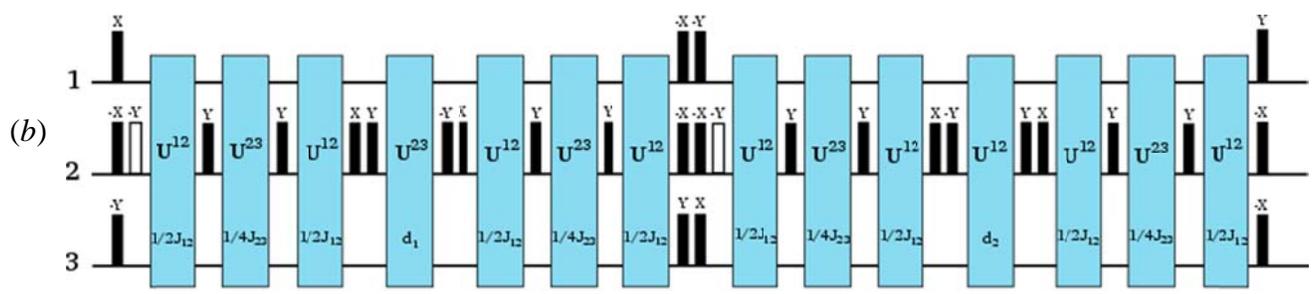

**FIG. 3.**

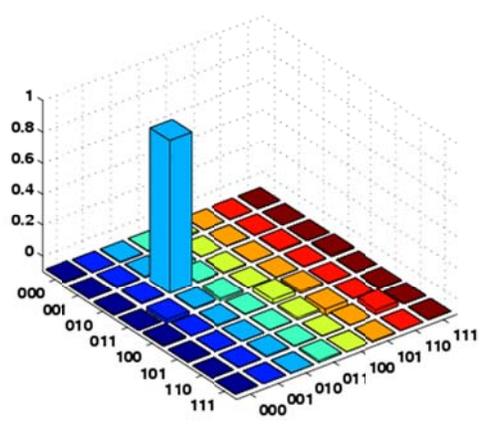

**FIG. 4.**

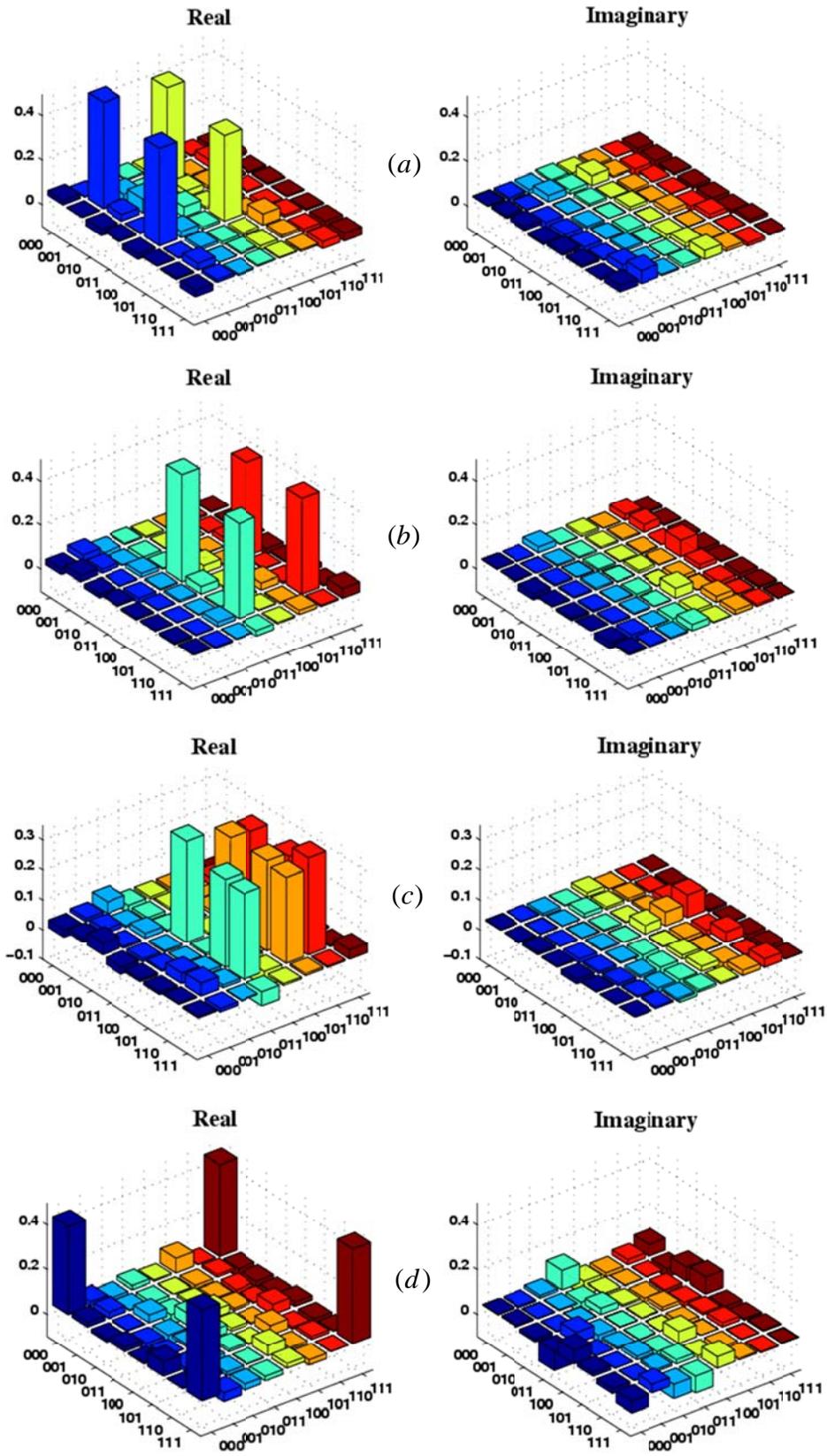

**FIG. 5.**

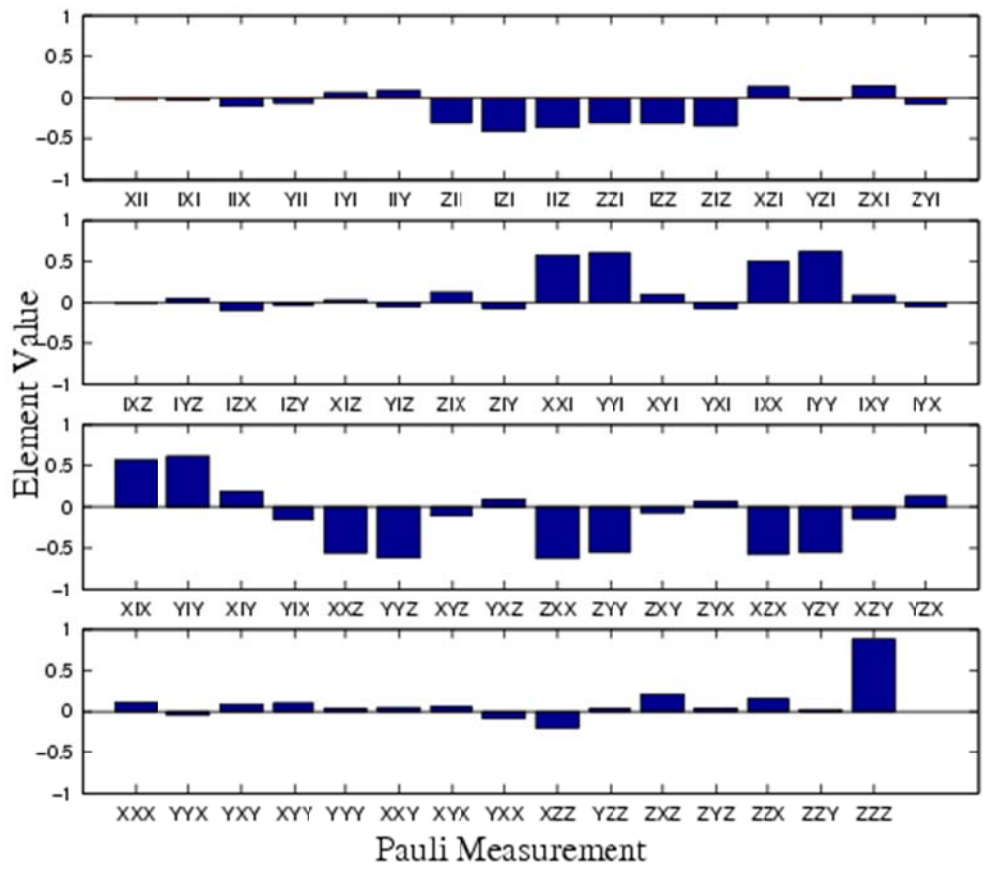

**FIG. 6.**

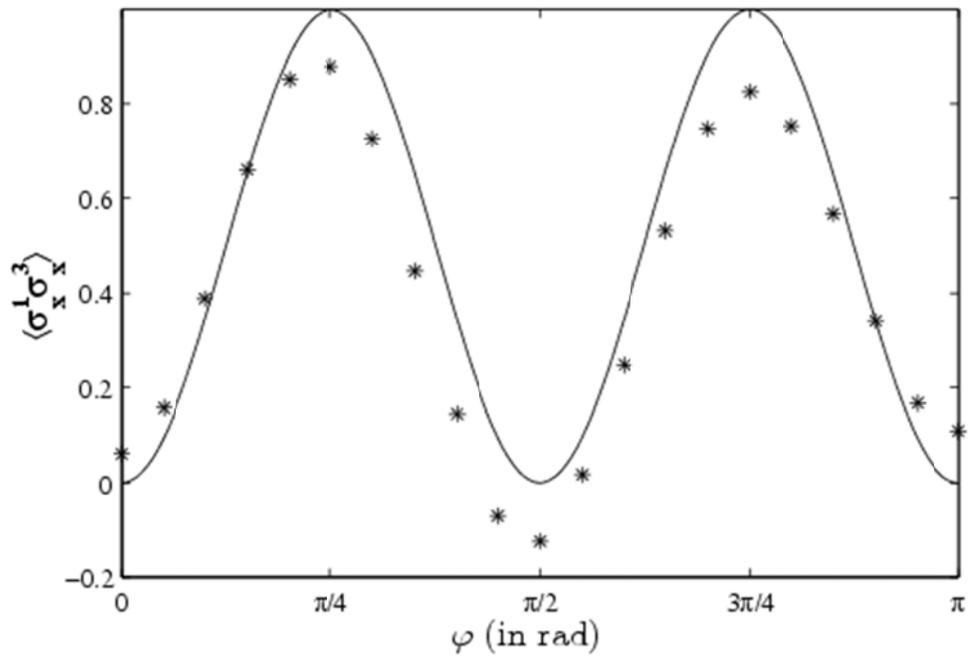

**FIG. 7.**

FIG. 1. (Color online) Entanglement dynamics in a 3-spin XY chain with nearest-neighbor interactions for different initial separable states, (a) $|010\rangle$ and $|101\rangle$, (b) $\frac{1}{\sqrt{2}}(|0\rangle + |1\rangle) \otimes \frac{1}{\sqrt{2}}(|0\rangle + |1\rangle) \otimes \frac{1}{\sqrt{2}}(|0\rangle + |1\rangle)$, (c) $|001\rangle$ and $|110\rangle$ (d) $|011\rangle$ and $|100\rangle$. Here, we use three different measures of entanglement, the bipartite entanglement, between the first and the last qubit $C_{13}$ (blue solid line), between the first and the second qubit $C_{12}$ (violet dashed line), the bipartite entanglement between the first and the remaining two qubits $C_{1(23)}$ (green dotted line), and the intrinsic three partite entanglement $C_{123}$ (red dash dotted line).

FIG. 2. Relevant properties of the molecule $^{13}$CHFBr$_2$: The left part shows the chemical structure of the molecule and the strengths (in Hz) of the J-couplings between the relevant nuclear spins. The right part shows the equilibrium spectra of $^1$H, $^{13}$C and $^{19}$F of $^{13}$CHFBr$_2$. The label on the top of each transition of a qubit represents the state of the other qubits in the transition.

FIG. 3. (Color online) The NMR pulse sequence for the implementation of the (a) PPS (b) unitary operator $U(t)$ of equation (16), which creates the entangled states. For the creation of Bell state on end qubits $d_1 = 1/2J_{23}$ and $d_2 = 1/2J_{12}$, for W-state $d_1 = 0.3041/J_{23}$ and $d_2 = 0.3041/J_{12}$, and for GHZ state $d_1 = 1/J_{23}$ and $d_2 = 1/J_{12}$. The empty rectangles represent 180º pulses and the black rectangles represent 90º pulses. The flip angle of the other pulses and the relevant phases of all the pulses are also written above them. The box with $U^{ij}$ represents the evolution under the J-coupling of spins $i$ and $j$ for a time period given in the box. The shaped pulses in (a) along G$_Z$ represent Z-gradients, which retain longitudinal magnetization at the end of the gradient.

FIG. 4. (Color Online) Real part of the experimentally determined density matrix for the initial state $|010\rangle$. The fidelity (equation (14)) was found to be $\approx 0.99$

FIG. 5. (Color online) The tomography of the real and imaginary parts of the experimentally obtained density matrices for the states (a) $\frac{1}{\sqrt{2}}(|001\rangle + |100\rangle)$, (b) $\frac{1}{\sqrt{2}}(|011\rangle + |110\rangle)$, (c) $\frac{1}{\sqrt{3}}(|101\rangle + |011\rangle + |110\rangle)$ (W-state), (d) $\frac{1}{\sqrt{2}}(|000\rangle + |111\rangle)$ (GHZ-state).

FIG. 6. Experimental Pauli set for the state $\frac{1}{\sqrt{3}}(|101\rangle + |011\rangle + |110\rangle)$. X, Y, Z represent Pauli operators and 'I' is single qubit identity operator. The trivial $\langle III \rangle = 1$ is not shown. The theoretically expected value for ZII, IZI, IIZ, ZZI, IZZ, ZIZ is -1/3, for XXI, YYI, IXX, IYY, XIX, YIY is 2/3, for XXZ, YYZ, ZXX, ZYY, XZX, YZY is -2/3, for ZZZ is 1, and all others are zero.

FIG. 7. The evolution of the two-body correlation $\langle \sigma_x^1 \sigma_x^3 \rangle$ of the end qubits under XY Hamiltonian for the initial state $|010\rangle$. The solid line indicates the ideal theoretical expectation and the points indicate the measured experimental results.